\documentclass[sigconf]{acmart}

\copyrightyear{2018} 
\acmYear{2018} 
\setcopyright{acmlicensed}
\acmConference[SIGIR '18]{The 41st International ACM SIGIR Conference on Research \& Development in Information Retrieval}{July 8--12, 2018}{Ann Arbor, MI, USA}
\acmBooktitle{SIGIR '18: The 41st International ACM SIGIR Conference on Research \& Development in Information Retrieval, July 8--12, 2018, Ann Arbor, MI, USA}
\acmPrice{15.00}
\acmDOI{10.1145/3209978.3209990}
\acmISBN{978-1-4503-5657-2/18/07}

\usepackage[subtle, wordspacing=normal, tracking=normal, bibnotes, charwidths=normal, leading, indent, lists, paragraphs=normal, mathspacing=normal, bibliography=normal]{savetrees}

\usepackage{bbm}
\usepackage{paralist}
\usepackage{booktabs} 
\usepackage{subfigure}
\usepackage{multirow}
\usepackage{array}
\usepackage{setspace}
\usepackage[skip=0pt]{caption}
\usepackage{enumitem}
\usepackage{acronym}

\acrodef{SERP}{search engine result page}
\acrodef{GUBM}{grid-based user browsing model}
\acrodef{NDCG}{normalized discounted cumulative gain}

\newcommand{\OurModel}{\acs{GUBM}}

\allowdisplaybreaks

\title{Constructing an Interaction Behavior Model\\ for Web Image Search}

\author{Xiaohui Xie}
\affiliation{%
\institution{Beijing National Research Center for Information Science and Technology, Department of Computer Science and Technology, Tsinghua University}
\city{Beijing}
\country{China}
}
\email{xiexh_thu@163.com}

\author{Jiaxin Mao}
\affiliation{%
\institution{Beijing National Research Center for Information Science and Technology, Department of Computer Science and Technology, Tsinghua University}
\city{Beijing}
\country{China}
}
\email{maojiaxin@gmail.com}

\author{Maarten de Rijke}
\orcid{0000-0002-1086-0202}
\affiliation{%
\institution{Informatics Institute, University of Amsterdam}
\city{Amsterdam}
\country{The Netherlands}
}
\email{derijke@uva.nl}

\author{Ruizhe Zhang}
\affiliation{%
\institution{Beijing National Research Center for Information Science and Technology, Department of Computer Science and Technology, Tsinghua University}
\city{Beijing}
\country{China}
}
\email{zhangrz15@mails.tsinghua.edu.cn}

\author{Min Zhang}
\affiliation{%
\institution{Beijing National Research Center for Information Science and Technology, Department of Computer Science and Technology, Tsinghua University}
\city{Beijing}
\country{China}
}
\email{z-m@tsinghua.edu.cn}

\author{Shaoping Ma}
\affiliation{%
\institution{Beijing National Research Center for Information Science and Technology, Department of Computer Science and Technology, Tsinghua University}
\city{Beijing}
\country{China}
}
\email{msp@tsinghua.edu.cn}

\renewcommand{\shortauthors}{X. Xie et al.}

\begin{abstract}
User interaction behavior is a valuable source of implicit relevance feedback. 
In Web image search a different type of search result presentation is used than in general Web search, which leads to different interaction mechanisms and user behavior. 
For example, image search results are self-contained, so that users do not need to click the results to view the landing page as in general Web search, which generates sparse click data. 
Also, two-dimensional result placement instead of a linear result list makes browsing behaviors more complex. 
Thus, it is hard to apply standard user behavior models (e.g., click models) developed for general Web search to Web image search.

In this paper, we conduct a comprehensive image search user behavior analysis using data from a lab-based user study as well as data from a commercial search log.
We then propose a novel interaction behavior model, called \ac{GUBM}, whose design is motivated by  observations from our data analysis. 
\ac{GUBM} can both capture users' interaction behavior, including cursor hovering, and alleviate position bias. 
The advantages of \ac{GUBM} are two-fold: \begin{inparaenum}[(1)] \item It is based on an unsupervised learning method and does not need manually annotated data for training. \item It is based on user interaction features on \acp{SERP} and is easily transferable to other scenarios that have a grid-based interface such as video search engines\end{inparaenum}.  
We conduct extensive experiments to test the performance of our model using a large-scale commercial image search log. 
Experimental results show that in terms of behavior prediction (perplexity), and topical relevance and image quality (\ac{NDCG}), \ac{GUBM} outperforms state-of-the-art baseline models as well as the original ranking. 
We make the implementation of \ac{GUBM} and related datasets publicly available for future studies.
\end{abstract}



\begin{document}

\maketitle

\section{Introduction}\label{section:introduction}
Interaction behavior data provides implicit but abundant user feedback and can be collected at very low cost~\citep{white-interactions-2016}. 
Thus, it has become a popular source for improving the performance of search engines. 
In particular, it has been successfully adopted to improve general Web search in result ranking~\cite{agichtein2006improving,o2016leveraging}, query auto-completion~\cite{li2014two,jiang2014learning}, query recommendation~\cite{cao2008context,wu2013learning}, optimizing presentations~\cite{wang2016beyond}, etc. 

\begin{figure}[h]
\centering
\includegraphics[clip,trim=5mm 0mm 0mm 0mm,width=\columnwidth]{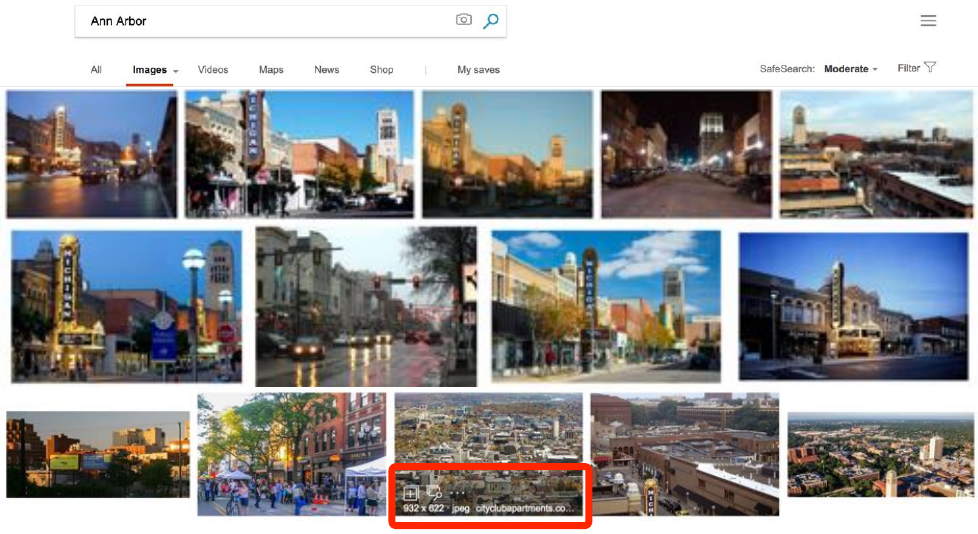}
\caption{An example \ac{SERP} produced by an image search engine. The red box shows the metadata of the image that is displaying when hovering over the corresponding position.}
\label{example_serps}
\end{figure}

Web image search results, however, are displayed in a markedly different way from general Web search engine results, which leads to different interaction mechanisms and differences in user behavior. 
Take the \acf{SERP} in Fig.~\ref{example_serps}, for example. 
An image search engine typically places results in a grid-based interface rather than a sequential result list. 
Users can view results not only in a vertical direction but also in a horizontal direction. 
Instead of a query-dependent summary of the landing page, the image snapshot is shown together with some metadata about the image; see the example highlighted with a red box in Fig.~\ref{example_serps}.
This type of metadata is usually only visible when the user hovers their cursor over the result. 
Also, while typically available on Web image search result pages, where users can view results by scrolling up and down without having to click on the ``next page'' button, pagination is usually not (explicitly) supported on general Web \acp{SERP}. 
Furthermore, image search results are self-contained, in the sense that users do not have to click the results to view the landing page as in general Web search, which leads to sparse click data. 
It has been demonstrated that the probability of an examined and relevant result being clicked is very low in Web image search~\cite{xie2017investigating}. 

Due to users' different and unique interactions with Web image search engines when compared with general Web search engines, it is hard to apply user behavior models that have been proven useful for general Web search to image search without  adaption. 
Take click models~\cite{chuklin-click-2015}, for example. 
They can alleviate user behavior bias and generate a reasonable estimation of result relevance in general Web search~\cite{chapelle2009dynamic,guo2009efficient,dupret2008user}. 
However, they tend to follow the sequential examination hypothesis~\cite{craswell2008experimental}. 
Also, the sparsity of clicks in image search generates a challenge to training these click-based models. 

In this paper, we conduct both a qualitative and quantitative analysis using data from a lab-based user study and data from a commercial search log to obtain a deeper understanding of user interactions with Web image search engines. 
These analyses then inform our modeling of Web image searchers. 
We find that users' unique and rich interactions are highly informative about their preferences. 
Specifically, we demonstrate that ``hovering'' over an image is highly correlated with users' examination behavior and result relevance. 
As image search interfaces elicit 8 to 10 times more hover interactions than clicks~\cite{park2015large}, cursor hovering could be a strong additional signal for relevance. 
Also, we find that examinations of images between two interaction signals (click and hover) usually follow one direction, but with possible skips. 
This observation shows that some of the assumptions used in previously proposed position-based models (e.g., the sequential examination assumption) are reasonable in a Web image search scenario, but only when restricted to the interval between two adjacent interaction signals. 
Motivated by these and other observations, we build a Web image search interaction model to capture user behavior. 
Compared with existing interaction models in image search, our model learns to rank results based on user behavior without incorporating text and visual contents. 
Also, the training process does not require manually labeled data.

To summarize, the main contributions of this work are as follows:

\begin{itemize}[nosep,leftmargin=8.5pt]
	\item By conducting a comprehensive analysis, using both a qualitative and a quantitative approach, using user study data and commercial search log data, we demonstrate the different user interactions with Web image search engines.  
	\item We propose a novel interaction behavior model named \acf{GUBM}. As this model is based on features of user interactions on \acp{SERP}, it can easily be transferred to other scenarios that have a two-dimensional interface such as video search engines.
	\item We conduct extensive experiments to test the performance of the proposed \ac{GUBM} model using a commercial search log data. The raw log data with image relevance and quality labels will be made publicly available (after the paper review process). The experimental results are promising in terms of behavior prediction, topical relevance, and image quality.  
\end{itemize}

\noindent%
We outline observations from our user study and query log analysis in Section~\ref{section:exploratoryanayses}. In Section~\ref{section:gridbaseduserbrowsingmodel}, we formally introduce \ac{GUBM}. 
We report on experiments using \ac{GUBM} and compare the results with results of existing models in Section~\ref{section:experimentsanddiscussions}. 
Section~\ref{section:related_work} reviews related work.  
Finally, conclusions and future work are discussed in Section~\ref{section:conclusionandfuturework}.


\section{EXPLORATORY ANALYSES}
\label{section:exploratoryanayses}
Although there also exist unique interactions with the image preview pages, the way to show preview pages varies with different search engines. For example, Google shows the preview page on the same page with the \ac{SERP} while Bing shows it on a new page. We leave the investigation of the preview pages as our future work.
In this paper, We perform exploratory analyses using two image search datasets to obtain a deeper understanding of user interactions with image search result pages.
One has been created using data collected in a lab-based user study and is publicly available~\cite{xie2017investigating}. 
In this user study, 40 participants have been recruited to complete 20 image search tasks which cover different query categories according to the Shatford-Panofsky approach~\cite{shatford1986analyzing}. 
Besides logged mouse activities, a Tobii eye-tracker has been used to record participants' examination behavior.
By using the built-in algorithms and all default parameters from Tobii Studio, the participants' fixation points were detected and the certain image being examined were recorded in the dataset.
Also, several professional assessors are recruited to provide relevance labels for query-image pairs. 
The second image search dataset that we use is a sampled of query log data from a commercial image search engine. 
The details about this dataset can be found in Section~\ref{subsection:datasets}.

\subsection{Hovering}
Interactions with an image result usually consist of mouse clicks, cursor hovering and user examination. 
Previous user behavior models developed for general Web search (e.g., click models) focus on mouse clicks. However, the number of clicks on image \acp{SERP} is small. In the commercial query log on which we conduct experiments (the second dataset), about 62.2\% of the sessions have no click. In contrast, there exist about 20 times more hover interactions than click. Also, \citet{park2015large} show that cursor hovering is strongly correlated with mouse clicks in image search. These findings motivate our first research question: 
\begin{description}
\item[\emph{\textbf{RQ1:}}] Can cursor hovering be an additional signal for relevance in image search scenarios? 
\end{description}
To answer this research question, we calculate a confusion matrix that shows the relation between hover actions (H) and relevance score (R) using user study data (the first dataset) and compare it to the confusion matrix in~\citep[Table 6]{xie2017investigating} which is based on clicks (C). 
The results are shown in Table~\ref{confusion_matrix}. 
\begin{table}[h]
\centering
\caption{Relation between click ($C$), hover ($H$) and relevance score ($R$) given images that are examined.}
\label{confusion_matrix}
\begin{tabular}{l|ll|ll}
\toprule
$E = 1$         & $C = 1$ & $C = 0$ & $H = 1$ & $H = 0$ \\ 
\midrule
Not relevant ($R=0$)    & 0.21\%       & 26.7\%      & 7.30\%       & 19.6\%      \\ 
Fairly relevant ($R=1$) & 0.41\%       & 6.94\%       & 2.61\%       & 4.74\%       \\ 
Very relevant ($R=2$)   & 4.84\%       & 60.9\%      & 25.7\%      & 40.0\%         \\ 
\bottomrule
\end{tabular}
\end{table}
Conditioned on the assumption that an image is examined, we can observe that the probability of a relevant result being hovered ($(2.61+25.7)\%/((2.61+4.74)+(25.7+40.0))\%=38.9\%$) is much higher than of being clicked ($(0.41+4.84)\%/((0.41+6.94)+(4.84+60.9))\%=7.18\%$). 
Also, the probability of a non-relevant result being hovered or clicked is low. 
Although the probability of hovering is slightly higher than the probability of a click on a non-relevant result, this potential noise can be compensated by the larger volume of hovering data (compared to click data). 
Thus, cursor hovering may be a useful signal for relevance in image search and it may help to overcome the sparseness of clicks. 

We modify the examination hypothesis~\cite{craswell2008experimental} from general Web search to our image search scenario by combining click and hover signals. 
We use \emph{interaction signal} (I) to indicate both \emph{click} (C) and \emph{hover} (H) and propose our first assumption: 
\begin{description}
\item[Assumption 1]
\textbf{-- Examination hypothesis in image search.} An image result being interacted with ($I_i=1$) accords with two independent conditions: it is examined ($E_i=1$) and it is relevant ($R_i=1$), which can be represented as:
\begin{equation}
\label{examination_hypothesis}
I_i=1 \leftrightarrow E_i=1 \mbox{ and }R_i=1.
\end{equation}
\end{description}

\subsection{Examinations between interactions}
\label{subsection:examinationsbetweeninteractions}
To construct a user interaction behavior model, it is essential to know how users examine results on a \ac{SERP}. 
Previous click models usually follow the sequential examination hypothesis~\cite{craswell2008experimental} according to which users examine results from top to bottom in a linear fashion. 
In Web image search engines, however, result placement is two-dimensional. Users can examine results not only in a vertical direction but also in a horizontal direction. 
\citet{xie2017investigating} analyze user examination behavior. 
One of their observations is that the probability of moving eyes horizontally is significantly higher than in other directions, which means that users tend to examine results within a single row before skipping to other rows. 
In this paper, we want to look deeper into users' examination behavior by considering the local context (that is, the interval between two adjacent interaction signals) to propose revised examination behavior assumptions for our interaction model. 
This motivates our second research question:

\begin{description}
\item[\emph{\textbf{RQ2:}}] How do users examine the image search results between two adjacent interaction signals?
\end{description}

\noindent%
We utilize the user study data to answer this research question. In an image search session, the interaction signals can be organized as a sequence $I = \langle I_1, I_2, \ldots,I_t, \ldots, I_T\rangle$ according to timestamp information, where  $1 \leq t\leq T$ and $T$ is the number of interaction signals in this search session. 
$I_t(r_t,c_t)$ records the position of the result being interacted with, where $r_t$ and $c_t$ represent the row number and column number, respectively. 
We separate the interaction sequence $I$ into adjacent pairs: $\langle I_0, I_1\rangle, \ldots, \langle I_n,I_{T+1}\rangle$ ($I_0(0,0)$ represents the search start of this session and $I_{T+1}(r_{max},c_{max})$ represents the search end of this session). 
In the user study data used in the exploratory analyses, $r_{max}$ is set to 20 (20 rows of results were reserved on the experimental result pages) and $c_{max}$ is the number of images in the final row. 

We then investigate the examination behavior between two adjacent interaction signals ($I_t$ and $I_{t+1}$) in the vertical direction and the horizontal direction, respectively. 

In the vertical direction, the relative spatial relations between two signals can be ``$\downarrow$'' ($r_t < r_{t+1}$) or ``$\uparrow$'' ($r_t > r_{t+1}$). 
In our data the proportion of ``$\downarrow$'' is 68.1\% and ``$\uparrow$'' is 31.9\%, which means that revisit behavior should be taken into account in image search scenarios. 
Intuitively, the examination direction would accord with the interaction direction while it is also possible that there exist some parts of direction change. 

\begin{figure}[h]
\centering
\includegraphics[clip,trim=1mm 0mm 0mm 0mm,width=\columnwidth]{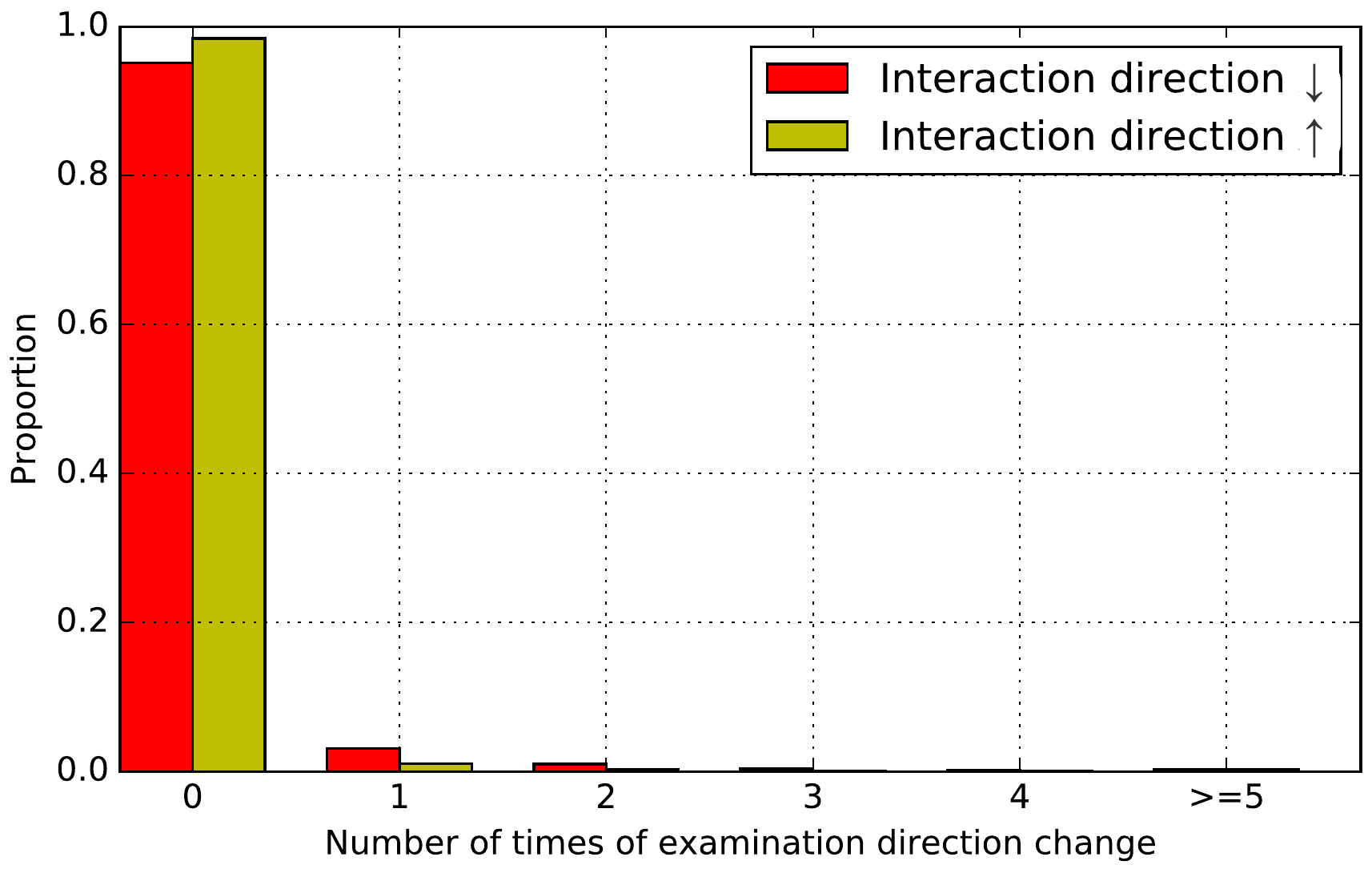}
\caption{Distribution of the fraction of examination direction changes for two types of adjacent interaction signals in vertical direction.}
\label{revisit_cnt_vertical}
\end{figure}

We count the number of times there is a change in examination direction for two types of adjacent interaction signals separately. 
From Fig.~\ref{revisit_cnt_vertical}, we can observe that in most cases (higher than 95\% for both types), users exhibit a sequential examination behavior following the same direction as the interaction direction without any direction change. 

\begin{figure}[h]
\centering
\includegraphics[clip,trim=2mm 0mm 0mm 0mm,width=\columnwidth]{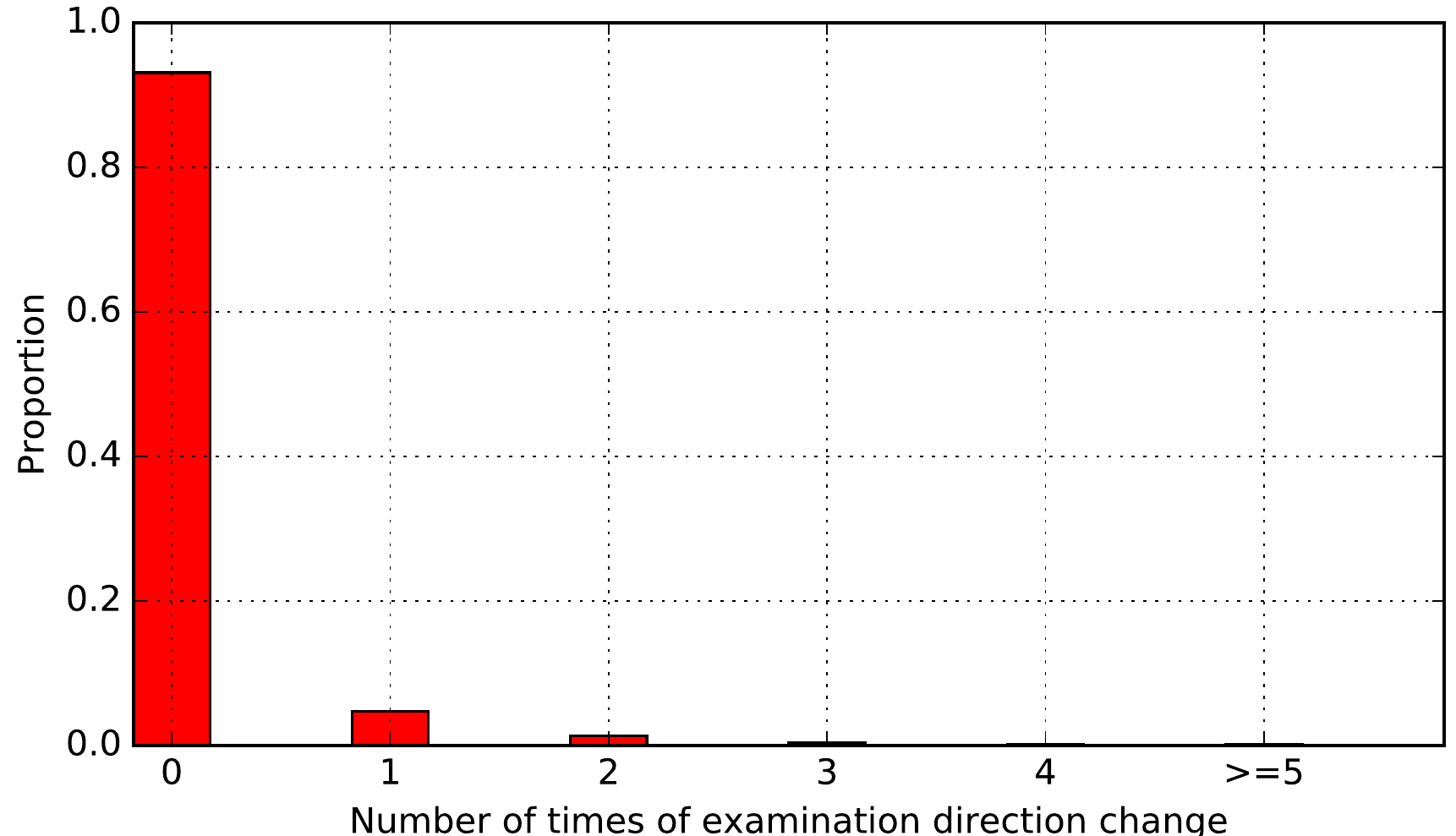}
\caption{Distribution of number of times of examination direction change in one row between two adjacent interaction signals.}
\label{revisit_cnt_horizontal}
\end{figure}

In the horizontal direction, we want to know how users examine results within a single row. We count the number of examination direction changes in each row (from row $r_t$ to row $r_{t+1}$) between adjacent interaction signals. Fig.~\ref{revisit_cnt_horizontal} displays the results in which more than 93\% examination sequences (within a single row) have no direction change. Thus, we can propose our second assumption: 
\begin{description}
\item[Assumption 2]
\textbf{-- Locally unidirectional examination assumption.} Between adjacent interaction signals, users tend to examine results in one direction without changes both in the vertical direction and in the same row. And the vertical examination direction is consistent with the vertical interaction direction. 
\end{description}

\subsection{Examination}
	\label{subsection:examination}
Determining the examination direction in a single row is difficult as there exist rows (about 40\% in our data) that receive fewer than two examination points. 
We predefine three types of in-row examination direction and calculate the proportion of sets of rows between adjacent interaction signal pairs that satisfy the definition of each type below. 
\begin{itemize}[nosep,leftmargin=8.5pt]
\item LtoR: Examination direction in every row follows a ``from left to right ($\rightarrow$)'' pattern: 60.5\% of sets of rows in our data.
\item RtoL: Examination direction in every row follows a ``from right to left ($\leftarrow$)'' pattern: 66.0\% of sets of rows  in our data.
\item Z-shape: Examination directions in adjacent rows are opposite: 93.2\% of all sets of rows in our data.
\end{itemize}
We assume that examination directions that receive fewer than two examination points can be from left to right ($\rightarrow$) or from right to left ($\leftarrow$), hence the percentages in the list just given add up to more than 100.

We can observe that the proportion of sequences of interaction signals that have a Z-shape is larger than the other two types. We can explain this finding using a perspective from~\cite{xie2017investigating}, viz.\ that users usually follow a ``nearby principle'' examination behavior in image search. 
In Section~\ref{section:experimentsanddiscussions} we will test the performance of models with different types of in-row examination direction.

As mentioned above at the beginning of Section~\ref{subsection:examination}, about 40\% of the rows receive less than two examination points; this means that users will not examine every result in the path between adjacent interaction signals. 
We further investigate the skipping behavior by answering the following research question:

\begin{description}
\item[\emph{\textbf{RQ3:}}] How far do users' eye gazes jump after examining the current result? 
\end{description}

\begin{figure}[h]
\centering
\includegraphics[clip,trim=2mm 0mm 0mm 0mm,width=\columnwidth]{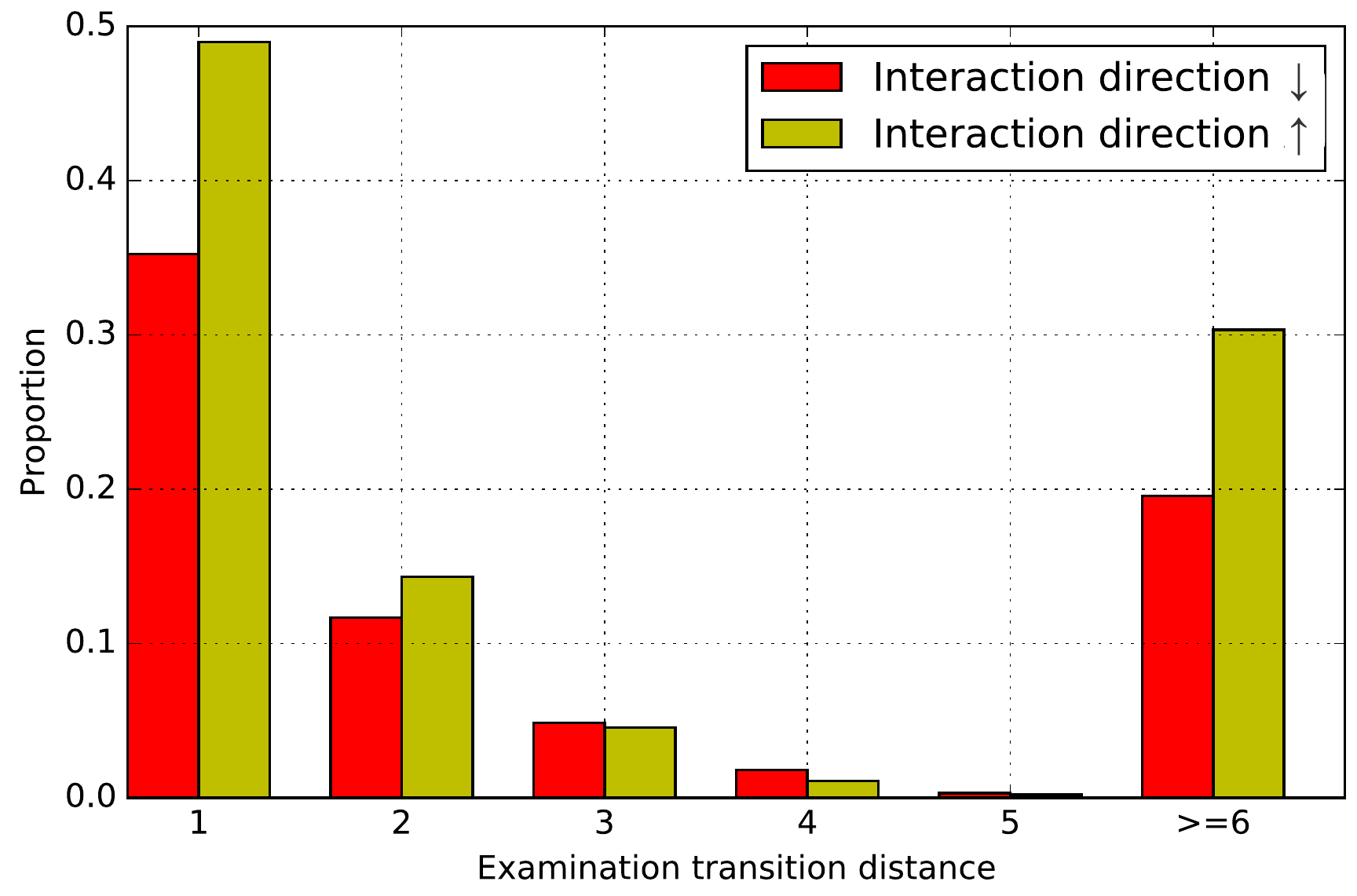}
\caption{Distribution of different examination transition distance  between two adjacent interaction signals.}
\label{skip_distance}
\end{figure}

\noindent%
We define the distance between two adjacent examined results as:
\begin{equation}
\label{img_distance}
d=\max(|r_{t+1}-r_{t}|,|c_{t+1}-c_{t}|),
\end{equation}
where $r$ refers to the row number and $c$ refers to the column number of examined results. 
We calculate the distance (as defined in Eq.~\ref{img_distance}) between all pairs of adjacent examinations in the path between two adjacent interaction signals; the results are shown in Fig.~\ref{skip_distance}. 
We can observe that more than 50\% of the adjacent examinations exhibit skipping behavior ($d > 1$).
The average examination transition distance for ``$\uparrow$'' (1.36) is significantly higher than for ``$\downarrow$'' (1.00) with $P$-value $< 0.001$. 
This may be caused by the fact that when users examine results from bottom to top (``$\uparrow$''), they will skip more results to re-examine a specific result. This brings us to our third assumption:
\begin{description}
\item[Assumption 3]
\textbf{-- Non one-by-one examination assumption.} When users examine the results between two adjacent interaction signals, they will skip several results and examine a result at some distance.
\end{description}

\begin{figure}[h]
\centering
\includegraphics[clip,trim=1mm 0mm 0mm 0mm,width=\columnwidth]{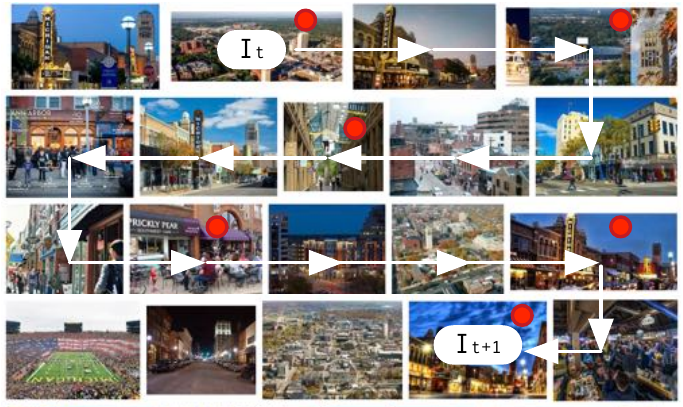}
\caption{A simulation of examination behavior on image \ac{SERP} between two adjacent interaction signals($I_t$ and $I_{t+1}$). We use white arrows to show the examination direction and red points to mark the examined results.}
\label{image_path}
\end{figure}

\noindent%
With answers to the above three research questions, RQ1--RQ3, we are able to obtain a clear picture of users' examination behavior between two adjacent interaction signals ($I_t$ and $I_{t+1}$). 
After interacting with an image result at row $r_t$ and column $c_t$ (click or hover over the image result),  a user will examine results with possible skips in the same row and will then move to the next row. 
The user will not revisit the previous rows until interacting with another image result at row $r_{t+1}$ and column $c_{t+1}$. 
The vertical direction of eye movements is consistent with the vertical interaction direction. 
Thus, given two adjacent interaction signals, we can consider examination behavior that takes place in between them to be in a linear way by joining adjacent rows together following a certain in-row examination direction. 
For example, by assuming a ``Z-shape'' interaction pattern, which satisfies most cases (93.2\%), we simulate examination behavior between interaction signals in Fig.~\ref{image_path}. 
We use white arrows to show the path of user examinations and red points to mark examined results. 

\section{Grid-based User Browsing Model}
\label{section:gridbaseduserbrowsingmodel}

In this paper, we aim to build an interaction behavior model based on interaction sequences that can be observed in large-scale commercial image search logs. 

\subsection{Model and hypotheses}
\label{subsection:modelandhypotheses}

Before introducing our model, named \acfi{GUBM}, we provide some definitions and notations. 

After a query $q$ has been issued to an image search engine, a \ac{SERP} is presented with a number of images that are organized in a grid-based manner. 
The result panel can be represented as a matrix: $\langle D_0,D_1,\ldots ,D_i,\ldots ,D_M \rangle $, where $D_i$ is the image sequence in the $i$-th row and can be represented as: $\langle d_{i0},d_{i1},\ldots,d_{ij},\ldots,d_{in_i} \rangle $ in which $d_{ij}$ is the image in the $i$-th row and $j$-th column and $n_i$ is the number of images in the $i$-th row. 
It should be noted that different rows may contain different numbers of images due to the different aspect ratios of the images. In Fig.~\ref{example_serps}, for example, $n_1$ equals 4 and $n_2$ equals 5. 

Guided by a search intent, the user begins to examine the \ac{SERP} and interacts with the results before abandoning the search. 
The interaction sequence can be recorded as $I = \langle I_1, I_2, \ldots,I_t, \ldots, I_T\rangle$, as introduced in Section~\ref{subsection:examinationsbetweeninteractions}. 
Given two adjacent interaction signals ($I_t(r_t,c_t)$ and $I_{t+1}(r_{t+1},c_{t+1})$), we define an \emph{image path} to be a sequence $P_{t,t+1}$ that can be generated based on Assumption~2 (locally unidirectional examination assumption) and a certain in-row examination direction; as illustrated in Fig.~\ref{image_path} (``Z-shape'' for the in-row examination direction), the image path $P_{t,t+1}$ can be represented as
\[
P_{t,t+1} = \langle d_{r_t,c_t}, d_{r_t,c_t+1},\dots,d_{r_{t+1},c_{t+1}+1},d_{r_{t+1},c_{t+1}} \rangle,
\]
where $d_{r_t,c_t}$ and $d_{r_{t+1},c_{t+1}}$ refer to the images being interacted by $I_t$ and $I_{t+1}$ respectively.
For every result ($d_{i,j}$) in the image path, we use a binary variable $E_{i,j}$ to indicate whether this result is being examined (1) or not (0). 
Because of Assumption~2 (locally unidirectional examination), we can simplify the subscript $i, j$, which indicates the $i$-th row and $j$-th column, to a single number $f(i,j)$ by considering the following mapping function, where $ED_i$ represents the examination direction in the $i$-th row which can be ``from left to right'' ($\rightarrow$) or ``from right to left'' ($\leftarrow$).
\begin{equation}
\label{mapping_function}
f(i,j)=\left\{
\begin{array}{ll}
\displaystyle
\sum_{r=0}^{i-1}{n_r}+j, & \text{if }{ (ED_i   \rightarrow) }\\[2ex]
\displaystyle
\sum_{r=0}^{i-1}{n_r}+n_i -j-1, & \text{if }{(ED_i   \leftarrow)}
\end{array} \right.
\end{equation}
Thus, the image path $P_{t,t+1}$ between $I_t$ and $I_{t+1}$ can be rewritten as $\langle d_m, \dots, d_i,\dots,d_n \rangle$ where $m=f(r_t,c_t)$ and $n=f(r_{t+1},c_{t+1})$. 

For each $d_i$ in the image path, we define the following binary variables to capture the user behavior over it: 
\begin{itemize}[nosep,leftmargin=8.5pt]
\item $\overline{I}_i$: the user interacted with the result by mouse click or cursor hovering. Note that $\overline{I}_i$ differs from the notation $I_t$ used to record the spatial position of interaction result.
\item $E_i$: the user examined the result.
\item $R_i$: the search result attracted the user as its content is relevant.
\end{itemize}
We are now in a position to describe our grid-based user browsing model.
According to the examination hypothesis in image search (Assumption~1 in Section~\ref{section:exploratoryanayses}), the probability of a result being interacted with can be described as Eq.~\ref{interaction_probability}:
\begin{equation}
P(\overline{I}_i = 1)  =  P(E_i=1)\cdot P(R_i = 1)
\label{interaction_probability}
\end{equation}
In this paper, we adopt the first-order hypothesis accepted in most user behavior models developed for general Web search~\cite{chapelle2009dynamic,dupret2008user}. 
The first-order hypothesis supposes that the click event at time $t+1$ is only determined by the click event at time $t$. We extend this hypothesis to construct our model by considering the adjacent interaction event:
\begin{equation}
P(I_{t+1} \mid I_t,\dots,I_1)  =  P(I_{t+1} \mid I_t)
\label{first_order_hypothesis}
\end{equation}
Next, Eq.~\ref{adjacent_interaction} depicts the user interaction in which we use the locally unidirectional examination assumption (Assumption~2) and a certain in-row examination direction to generate an image path between two adjacent interaction signals:
\begin{eqnarray}
\lefteqn{P(I_{t+1}=(r_{t+1},c_{t+1}) \mid I_t=(r_t,c_t)) {}}
\label{adjacent_interaction} \\
&=& P(I_{t+1} =f(r_{t+1},c_{t+1})=n \mid I_t=f(r_t,c_t)=m) \nonumber \\
&=& P(\overline{I}_m=1,\dots,\overline{I}_i = 0,\dots,\overline{I}_n = 1)\nonumber
\end{eqnarray}
By considering the image path between $I_t$ and $I_{t+1}$, the probability of an image result being examined depends on the spatial position of this result in the path. 
We use an internal parameter $\gamma_{imn}$ to depict the examination probability while the vertical examination direction along the path can be either $\downarrow$ $(m \leq n)$ or $\uparrow$ $(n \leq m)$: 
\begin{eqnarray}
\lefteqn{P(E_i=1 \mid I_{t}=m,I_{t+1}=n) }
\label{exam_probability} \\
& = & \left\{
\begin{array}{ll}
\gamma_{imn} & \text{if } m \leq i \leq n\text{ or }n \leq i \leq m  \\
0 & \text{otherwise.}
\end{array}
\right.\nonumber 
\end{eqnarray}
Examination may also be affected by other image content features such as saliency or edge density~\cite{xie2017investigating}. 
In this paper, however, we only incorporate position information into our model; other features will be discussed in possible directions for future work.

Next, another internal parameter used is $\alpha_{uq}$, which represents the relevance level for a query-image pair, where $u$ is the image result for the specific query $q$ that is at position $i$ in the image path:
\begin{eqnarray}
P(R_i=1) & = &\alpha_{uq}
\label{relevance_probability}
\end{eqnarray}
The task here is to infer $\gamma$ and $\alpha$ from observable data (i.e., interaction sequence $I$). 
We assume that these two parameters are user independent so as to simplify the model. 

In summary, Eqn.~\ref{interaction_probability}--\ref{relevance_probability} above encode the assumptions used by \ac{GUBM} to simulate user behavior on an image \ac{SERP}.

\begin{figure}[h]
\centering
\includegraphics[clip,trim=1mm 0mm 0mm 0mm,width=\columnwidth]{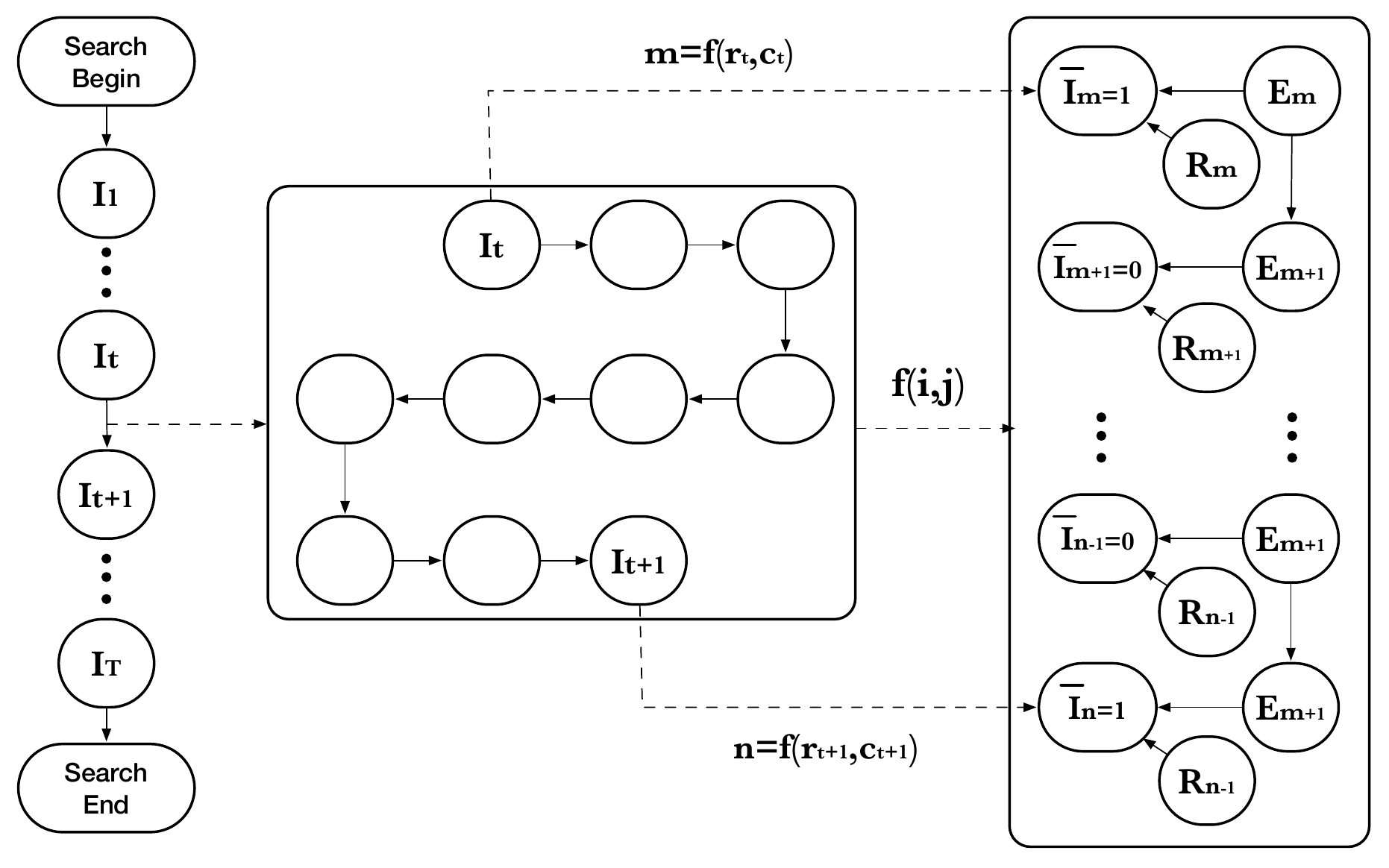}
\caption{Graphical representation of grid-based user browsing model (\ac{GUBM}).}
\label{graphical_representation}
\end{figure}

Fig.~\ref{graphical_representation} provides a graphical representation of \ac{GUBM}. 
From search start to search end, an interaction sequence $\langle I_1,\dots,I_T \rangle $ can be observed. 
Considering an adjacent interaction signals, the image path between them can be obtained by predefined in-row examination direction (i.e., ``Z-shape''). 
Then, based on the locally unidirectional examination assumption (Assumption~2), we can investigate user examinations between two adjacent interaction signals in an approximately linear way, which enables us to borrow some ideas from click models in general Web search. 
Specifically, the assumptions derived from data analysis share some perspective with partially sequential click model (PSCM)~\cite{wang2015incorporating}.

\subsection{Model inference for \ac{GUBM}}
\label{subsection:modelinferenceforGUBM}

In this paper, we use the expectaction-maximization (EM) algorithm~\cite{dempster1977maximum} to infer the internal parameters $\gamma$ and $\alpha$ from search logs. 
In \ac{GUBM}, the observation is the interaction sequence ($I$) while the hidden variables are the query-image relevance level ($R$) and examination behavior ($E$). 
Given internal parameter $\theta=\{{\gamma,\alpha}\}$, the marginal likelihood can be represented as Eq.~\ref{marginal_likelihood} according to Eq.~\ref{first_order_hypothesis}.
\begin{equation}
P(I,E,R \mid \theta) = \prod_{t=0}^{T}P(I_{t+1},E,R \mid I_t,\theta) 
\label{marginal_likelihood}
\end{equation}
We use $s$ to denote a specific query-session and $S$ to denote all sessions in the search logs; $I^s$ refers to the interaction sequence in session $s$. 
The conditional expected log-likelihood (Q-function) can then be written as:
\begin{equation}
\label{q_function}
Q=\sum_{s\in S} E_{E,R \mid I^s,\theta}[\log P(I^s,E,R \mid \theta)]
\end{equation}
Thus, in iteration round $v$ of the EM algorithm, the Q-function of $\alpha_{uq}$ and $\gamma_{imn}$ can be formulated as:
\begin{alignat}{2}
Q_{\alpha_{uq}} &= \sum_{s\in S}\sum_{(I_t,I_t+1) \in I^s} \mathbbm{1}_{i \neq n } \cdot \frac{(1-\alpha_{uq}^v )\log(1- \alpha_{uq})}{1-\alpha_{uq}^v\gamma_{imn}^v} + {} \label{q_function_alpha}
\\
& \qquad \mathbbm{1}_{i \neq n } \cdot \frac{\alpha_{uq}^v(1-\gamma_{imn}^v ) \log(\alpha_{uq})}{1-\alpha_{uq}^v\gamma_{imn}^v} + \mathbbm{1}_{i = n} \cdot \log(\alpha_{uq})
\nonumber
\\[1.1ex]
Q_{\gamma_{imn}} & =  \sum_{s\in S}\sum_{(I_t,I_t+1) \in I^s} \mathbbm{1}_{i \neq n } \cdot \frac{(1-\gamma_{imn}^v )\log(1- \gamma_{imn})}{1-\alpha_{uq}^v\gamma_{imn}^v} + {} \label{q_function_gamma}
\\
&\qquad  \mathbbm{1}_{i \neq n } \cdot \frac{\gamma_{imn}^v(1-\alpha_{uq}^v ) \log(\gamma_{imn})}{1-\alpha_{uq}^v\gamma_{imn}^v} + \mathbbm{1}_{i = n} \cdot \log(\gamma_{imn}),
\nonumber
\end{alignat}
where $u$ is shown in the image path $\langle d_m,\dots,d_n\rangle$ with subscript $i$ between interactions $I_t$ and $I_{t+1}$ in session $s$. 
Furthermore, $\theta^v=\{{\alpha^v,\gamma^v}\}$ are the internal parameters in iteration round $v$. 
And $\mathbbm{1}$ is the indicator function.

To maximize the Q-function, we take the derivate of Eq.~\ref{q_function_alpha} and Eq.~\ref{q_function_gamma} with respect to the parameters $\alpha_{uq}$ and $\gamma_{imn}$, respectively. Then, we can obtain update functions for $\alpha_{uq}$ and $\gamma_{imn}$ as:
\begin{eqnarray}
aF_1^{v} &= \mathbbm{1}_{i \neq n } \cdot \frac{1-\alpha_{uq}^v}{1-\alpha_{uq}^v\gamma_{imn}^v} 
\nonumber \\
aF_2^{v} &= \mathbbm{1}_{i \neq n } \cdot \frac{\alpha_{uq}^v(1-\gamma_{imn}^v)}{1-\alpha_{uq}^v\gamma_{imn}^v} 
\label{update_function_alpha}\\
aF_3^{v} &= \mathbbm{1}_{i = n } \cdot 1 \nonumber \\
\alpha_{uq}^{v+1} &=\sum_{s\in S}\sum_{(I_t,I_t+1) \in I^s} \frac{aF_2^{v}+aF_3^{v}}{aF_1^{v}+aF_2^{v}+aF_3^{v}}
\nonumber 
\end{eqnarray}
and
\begin{eqnarray}
gF_1^{v} & =& \mathbbm{1}_{i \neq n } \cdot \frac{1-\gamma_{imn}^v}{1-\alpha_{uq}^v\gamma_{imn}^v} 
\nonumber\\
gF_2^{v} & = & \mathbbm{1}_{i \neq n } \cdot \frac{\gamma_{imn}^v(1-\alpha_{uq}^v)}{1-\alpha_{uq}^v\gamma_{imn}^v} 
\label{update_function_gamma}\\
gF_3^{v} & = & \mathbbm{1}_{i = n } \cdot 1 
\nonumber\\
\gamma_{imn}^{v+1} & = & \sum_{s\in S}\sum_{(I_t,I_t+1) \in I^s} \frac{gF_2^{v}+gF_3^{v}}{gF_1^{v}+gF_2^{v}+gF_3^{v}}
\nonumber
\end{eqnarray}

\noindent%
In this paper, we set the initial value of $\alpha_{uq}$ and $\gamma_{imn}$ to 0.5 and the number of iterations to 40. 
For each session $s$, we investigate the top 100 images and the interaction sequence over them as the proportion of queries with around 100 results being shown is much higher than other queries in our dataset from a commercial image search engine. 
(In commercial image search engines, a result is loaded and shown only when users scroll to the area the result belong to, and the number of results being loaded and shown will be recorded in search logs.)


\section{EXPERIMENTS AND DISCUSSIONS}
\label{section:experimentsanddiscussions}

We evaluate the proposed grid-based user browsing model, \OurModel{}, using search logs from a commercial image search engine. As our model can estimate not only the query-image relevance level but also the interaction (click and hover) probability of image results based on Eq.~\ref{exam_probability}, we conduct extensive experiments including topical relevance and image quality estimation (in terms of normalized discounted cumulative gain (\ac{NDCG})) as well as behavior prediction (in terms of perplexity). 

\subsection{Datasets}
\label{subsection:datasets}

The experimental dataset is randomly sampled from a search log in October 2017. 
We discard queries with fewer than 10 search sessions to make sure that the proposed interaction model and baseline models can capture enough information, as in~\cite{dupret2008user}. 
Also, for each query, we reserve 1,000 query sessions to prevent a small number of queries from dominating the data. 
The details about the dataset can be found in Table~\ref{experiment_dataset}. 
We split all query sessions into training and test sets at a ratio of 7:3, following previous publications~\cite{xu2012incorporating,o2016leveraging,chen2012beyond}. 

\begin{table}[]
\centering
\caption{Description of experiment dataset (``\#'' refers to ``number of''). $H$ ($C$) sessions are sessions that have at least one hover ($H$) (click ($C$)) action.}
\label{experiment_dataset}
\begin{tabular}{cccc}
\toprule
\#Distinct queries & \#Sessions & \#$H$ Sessions & \#$C$ Sessions \\ \midrule
16,194              & 476,586     & 466,208           & 179,410           \\ 
\bottomrule
\end{tabular}
\end{table}

The relevance for Web image search consists of two facets: topical relevance and image quality~\cite{geng2011role,o2016leveraging}; we gather annotations for these two facets respectively first and then incorporate them together as the same way in~\cite{o2016leveraging}. 

For topical relevance, we measure the relation between subjects in a text query and visual subjects in the image. We gather judgment on the following 3-point scale:

\begin{itemize}[nosep,leftmargin=8.5pt]
\item Not relevant (0): The visual subjects fail to match the subjects in the text query. (Example: the query is ``Bicycle'' and the main object in the image is ``Car''). 
\item Fairly relevant (1): The visual subjects partly match the subjects in the text query or vice versa, which can mean three things: (1)~The query contains two or more objects while the image only describe part of them (Example: the query is ``Cat and mouse'' and the image only depicts ``Cat''). (2)~The image contains more objects than the query can fully describe. (Example: the query is ``Cat'' and the main objects in the image include both ``Cat'' and ``Mouse.'')  (3)~Although the objects are matched between the two modalities, their modifiers are different (Example: the query is ``Red Ferrari'' and the image is about ``Black Ferrari''). 
\item Very relevant (2): The visual subjects perfectly match the subjects in text query.
\end{itemize}

\noindent%
We measure image quality according to framework (the position of the main objects in the image), clarity, watermark, color and brightness, and the judgments are gathered on the following 5-point scale:

\begin{itemize}[nosep,leftmargin=8.5pt]
\item Bad (0): Badly framed, big and obvious watermark, low clarity.
\item Fair (1): Fairly framed, small but obvious watermark, slightly blurred.
\item Good (2): Well framed, small and not easily perceived watermark, low value for download or image collections.
\item Excellent (3): Nicely framed, no watermark, fairly attractive and appealing.
\item Perfect (4): Without aesthetic flaws, very attractive and appealing, high artistic value.
\end{itemize}

\noindent%
For each query-image pair topical relevance judgment or image quality judgment, at least three editors are recruited to provide the annotation based on the above instructions. The Fleiss Kappa scores~\cite{fleiss1971measuring} among annotators are higher than 0.5 (substantial agreement) for both two tasks. 

The final 5-scale relevance score can be obtained by the principle~\cite{o2016leveraging} that the relevance score of an image equals the topical relevance score when the topical relevance score is 0 or 1 and equals the image quality score when the topical relevance score is 2. 

We sample 500 distinct queries from the dataset; after filtering out pornographic queries, 448 distinct queries and around 50,000 images are annotated. The distribution of different relevance scores is 9.33\% (0), 17.0\% (1), 18.0\% (2), 54.3\% (3), and 1.43\% (4).

\subsection{Evaluation of behavior prediction}
\label{subsection:evaluationofbehaviorprediction}
To measure the effectiveness of an interaction behavior model on behavior prediction, an often used metric is perplexity~\cite{dupret2008user}. 
Previous work on click models applies perplexity to test how well a model can predict clicks~\cite{wang2015incorporating,liu2017time}. 
In this paper, we use perplexity to test the ability of \ac{GUBM} on predicting interaction behavior (click and hover together) and validate the assumptions proposed in Section~\ref{section:exploratoryanayses} by comparing \ac{GUBM} to other user behavior models. The formulation of perplexity can be written as:
\begin{equation}
\label{perplexity}
p_r(M)= 2^{-\frac{1}{|S|} \sum_{s \in S }(\overline{I}_r^s \log_2 q_r^s + (1 - \overline{I}_r^s) \log_2(1- q_r^s ) ) },
\end{equation}
where $M$ refers to the model we evaluate and $q_r^s$ is the probability of the image result, at rank $r$ in  session $s$, being interacted with. 
Please note that $r$ is mapped from tuple position (row, column) to a numerical value using Eq.~\ref{mapping_function} with ``LtoR'' as the in-row examination direction (commercial image search engines usually place results from left to right in each row). 
A smaller perplexity value indicates better prediction performance; the ideal perplexity value is 1. 
The improvement of perplexity $P^A$ over $P^B$ can be calculated by $\frac{P^B-P^A}{P^B-1}$.

In this paper, we compare the following models to validate the proposed assumptions (specially, Assumption 2 and~3):

\begin{itemize}[nosep,leftmargin=8.5pt]
\item \ac{GUBM}-LtoR, \ac{GUBM}-RtoL, \ac{GUBM}-Zshape:  These three models are \ac{GUBM} with different in-row examination directions (see Section~\ref{subsection:examination}). By comparing them, we want to investigate which type of in-row examination direction fits the user browsing behavior better.
\item UBM \cite{dupret2008user}: A classic position-based model. Although this model also allows for users' skipping behavior, it makes the sequential examination assumption, which assumes that users will not revisit previous results. We transfer two-dimen\-sional result panels to result lists by using Eq.~\ref{mapping_function}, following a ``Z-shape'' in-row examination direction to instantiate UBM for a grid-based interface.
\item THCM~\cite{xu2012incorporating}: A sequence-based model that allows for users' revisit behavior. However, it requires that user examine results one-by-one. We adapt it to the image search scenario by the same method that we used for UBM above. 
\item POM~\cite{wang2010inferring}: A flexible model that allows arbitrary examination behavior between two adjacent signals without constraint. 
\end{itemize}

\begin{figure}[h]
\centering
\includegraphics[clip,trim=2mm 0mm 0mm 0mm,width=\columnwidth]{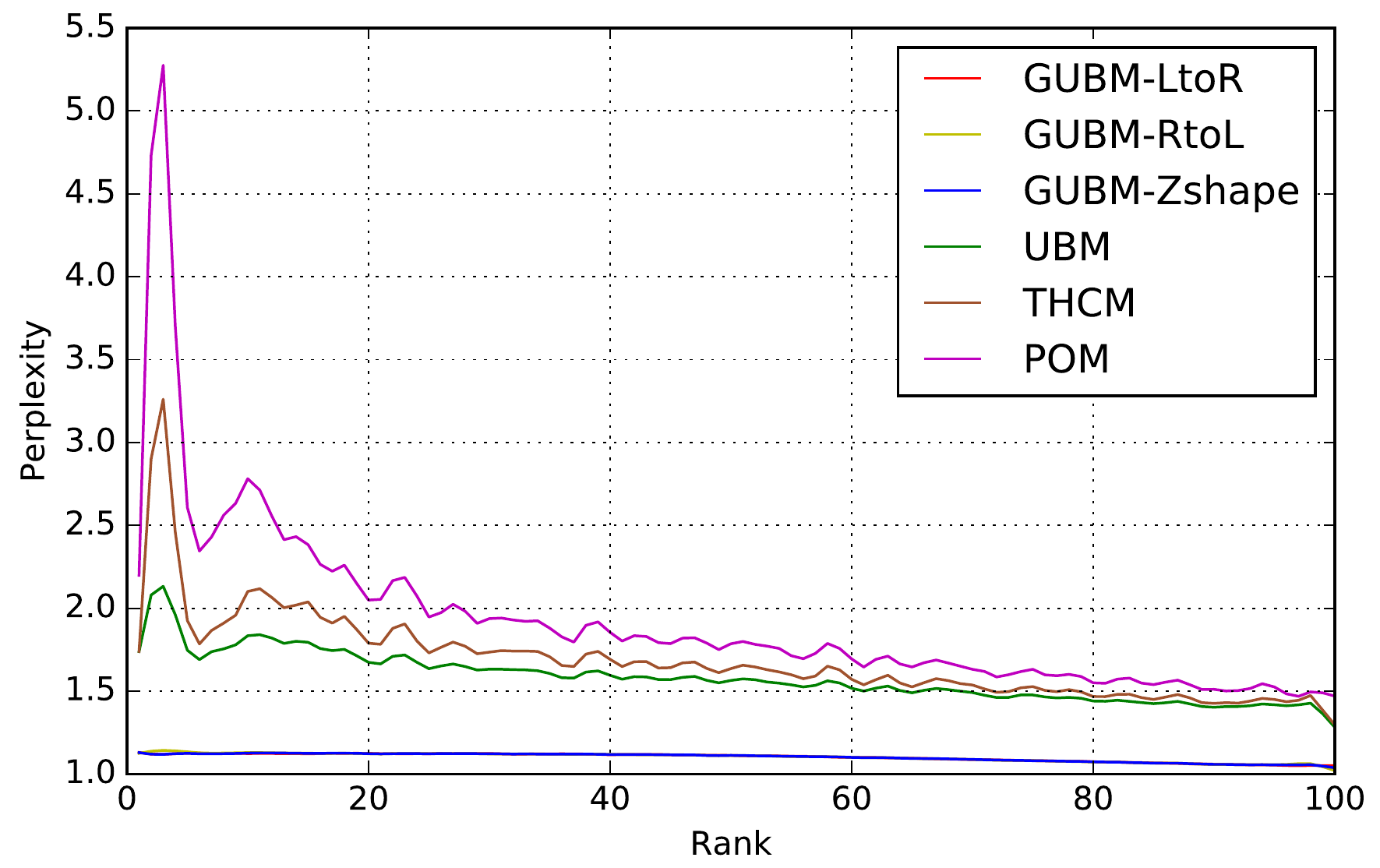}
\caption{Interaction perplexities of different ranks of different models.}
\label{perplexity_plot}
\end{figure}

\begin{table}[]
\centering
\caption{Overall interaction perplexity (average perplexities at different ranks) of each model (all improvements are significant with $p$-value $< 0.001$. )}
\label{overall_perplexity}
\begin{tabular}{lrr}
\toprule
Model       & Overall Perplexity & GUBM-Zshape Impr. \\ \midrule
UBM         & 1.5806             & 82.6\%            \\ 
THCM        & 1.6876             & 85.3\%            \\ 
POM         & 1.9250             & 89.0\%            \\ 
GUBM-RtoL   & 1.1017             & --                 \\ 
GUBM-LtoR   & 1.1011             & --                 \\ 
GUBM-Zshape & 1.1010             & --                 \\ 
\bottomrule
\end{tabular}
\end{table}

In Fig.~\ref{perplexity_plot} we plot the perplexities at different ranks (1--100) for each model; and in Table~\ref{overall_perplexity} we show the overall perplexity obtained by averaging the perplexities at different ranks. 

From Fig.~\ref{perplexity_plot}, we can observe that there is no visible difference between \ac{GUBM}-LtoR, \ac{GUBM}-RtoL and \ac{GUBM}-Zshape, although from Table~\ref{overall_perplexity}, the overall perplexity of \ac{GUBM}-Zshape is slightly better. 
In an image search scenario, users will follow unidirectional in-row examination behavior and the in-row direction can be more flexible than the predefined directions introduced in Section~\ref{subsection:examination}. 
Intuitively, a ``nearby principle'' will be a better fit for users' examination behavior between adjacent rows. 
By comparing \ac{GUBM}-Zshape to other popular user behavior models (UBM, THCM and POM), we find that the performance of \ac{GUBM}-Zshape is significantly better on behavior prediction, at different ranks and in overall comparison. 
The improvements show that the assumptions proposed in Section~\ref{section:exploratoryanayses} are much closer to practical user behavior than the assumptions underlying the competing models. 
We can conclude that considering users revisit behavior is useful for behavior modeling (\ac{GUBM} vs.\ UBM) and users will display a skipping behavior rather than examine results one-by-one (\ac{GUBM} vs.\ THCM, Assumption 3). 
Also, the locally unidirectional assumption fits user behavior well (Assumption~2). 
Although considering examination behavior between adjacent interaction signals is much more flexible (POM), it may give the model too much freedom for it  to be correctly parameterized in practice.

\begin{table}[]
\centering
\caption{Relevance (topical relevance+image quality) estimation performance in terms of NDCG@5, 10, 15 and 20 (448 distinct queries). ** (*): The difference is significant between the baseline model and \ac{GUBM} with $p$-value < $0.01$ ($0.05$). }
\label{NDCG_table}
\begin{tabular}{lccccc}
\toprule
Model   & Original ranking &UBM & GUBM-C & GUBM & ~  \\ \midrule
NDCG@5  & 0.9165\rlap{**}           &0.9231\rlap{*} & 0.9242\rlap{*} & \textbf{0.9349} \\ 
NDCG@10 & 0.9078\rlap{**}           &0.9084\rlap{**} & 0.9090\rlap{**} & \textbf{0.9252} \\
NDCG@15 & 0.9065\rlap{*}           &0.9066\rlap{*} & 0.9063\rlap{*} & \textbf{0.9179} \\ 
NDCG@20 & 0.9049\rlap{*}           &0.9042\rlap{*} & 0.9047\rlap{*} & \textbf{0.9159} \\ \bottomrule
\end{tabular}
\end{table}

\subsection{Evaluation of relevance estimation}

An important goal of our interaction behavior model is to improve the search ranking of image search engines. As our model can provide relevance estimation of a query-image pair, we can rerank the original ranking according to the estimated score ($\alpha_{uq}$). 
In this paper, we apply normalized discounted cumulative gain (NDCG)~\cite{jarvelin2002cumulated} to measure the performance of a given ranking. 
For a ranked list of images, the DCG score is defined as:
\begin{equation}
\label{NDCG}
DCG@d = \sum_{i=1}^{d} \frac{r_i}{\log_2(i+1)},
\end{equation}
where $r_i$ is the relevance score at position $i$ and $d$ is the depth of this ranked list of images. 
Then the NDCG@d can be obtained by normalizing DCG@d using ideal DCG@d which measures the perfect ranking. 
We utilize the annotated data introduced in Section~\ref{subsection:datasets} and compare the performance of \ac{GUBM} (with ``Z-shape'' in-row examination direction) on relevance (topical relevance+image quality) estimation with UBM introduced in Section~\ref{subsection:evaluationofbehaviorprediction} and the following two baselines:

\begin{itemize}[nosep,leftmargin=8.5pt]
\item Original ranking: The original search result ranking returned after issuing a certain query by the commercial search engine from which our data was collected. To be noted, the original ranking is recorded at the same time as our dataset.
\item \ac{GUBM}-C: A model trained on the full dataset in which we remove hover information in all sessions while we preserve click information. By comparing GUBM-Zshape to this model, we want to investigate whether hover information plays an important role in improving image search result ranking.
\end{itemize}

We compare the proposed \ac{GUBM}-Zshape with these two baselines in terms of NDCG@5, 10, 15 and 20. 
The results are listed in Table~\ref{NDCG_table}. 
Our model is significantly better than UBM which indicates that the underlying assumptions of our model are more appropriate in an image search scenario.
We also observe that our model achieves significant improvements over both \ac{GUBM}-C and the original ranking on all metrics. 
The results indicate that hover can be a strong additional signal for relevance and Assumption~1 (Examination hypothesis in image search) can be useful for behavior modeling in an image search scenario. 

We examine different cut-offs and find that the ranking performance of \ac{GUBM}-C is almost the same as the original ranking in terms of NDCG@15 and NDCG@20. 
When the depth of the ranked list of images is reduced (to 10 or even 5), the average performance of \ac{GUBM}-C improves and the improvement itself increases too, although no significant differences are observed. 
This phenomenon may be caused by the fact that users tend to click the top results~\cite{xie2017investigating}, so that click information is relatively abundant at the top positions, which may guide the model in the right direction. 
However, when considering more results (i.e., top 15 and 20), the sparsity of clicks may result in a smaller and unobservable improvement. 

\begin{table}[h]
\centering
\caption{Topical relevance ($TR$) and Image quality ($I$) estimation performance of GUBM-Zshape in terms of NDCG@5 and 10 (448 distinct queries). $\ddagger$ ($\dagger$): better than the original ranking with $p$-value < $0.01$ ($0.05$). }
\label{topical_relevance_NDCG}
\begin{tabular}{lcccc}
\toprule
Model            & TR@5             & TR@10           & I@5               & I@10             \\ 
\midrule
Original ranking & 0.9713           & 0.9711          & 0.9396            & 0.9324           \\ 
GUBM-Zshape             & \textbf{0.9783\rlap{$\dagger$}} & \textbf{0.9749} & \textbf{0.9530\rlap{$\ddagger$}} & \textbf{0.9442\rlap{$\dagger$}} \\ \bottomrule
\end{tabular}
\end{table}

We also report on the performance of GUBM-Zshape in terms of topical relevance and image quality estimation separately in Table~\ref{topical_relevance_NDCG} by comparing \ac{GUBM}-Zshape with the original ranking. 
The results indicate that \ac{GUBM}-Zshape can improve the ranking both in topical relevance and image quality and that the improvement in image quality ($I$) is larger than in topical relevance ($TR$). Also, we should note that the average improvement is not significant in terms of TR@10. The reason for this may be two-fold: (1)~as we compute the NDCG of the top ranking, the score is already very high enough (the average level is higher than 0.97) so that the room for improvement is limited; (2)~user interaction behavior in an image search scenario may have a closer correlation with image quality, thus confirming~\cite{geng2011role}. 

\subsection{Case studies}

\begin{figure}[t]
\centering
\includegraphics[clip,trim=2mm 0mm 0mm 0mm,width=\columnwidth]{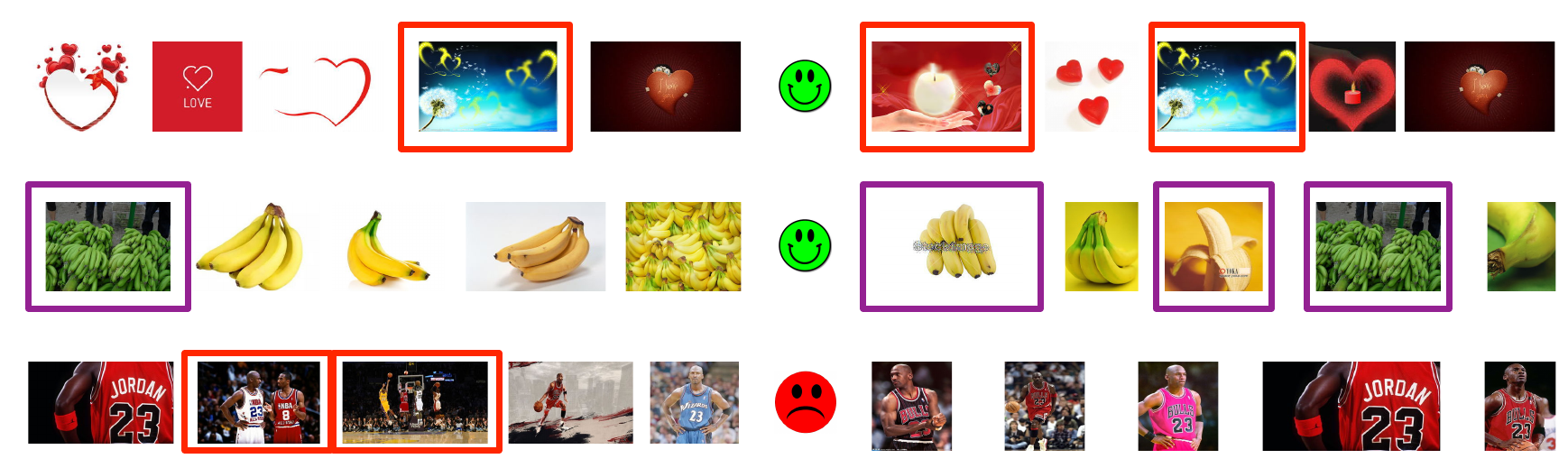}
\caption{Qualitative comparison of results re-ranked by \ac{GUBM}-Zshape (left) and results from the original ranking (right). \ac{GUBM} outperforms the original ranking in the top 2 cases (``loving heart'' and ``banana'') and loses in the last case (``Michael Jordan''). (Red boxes mark images with a low topical relevance level given the query. Purple boxes mark images with poor image quality.)}
\label{case_study}
\end{figure}

Besides evaluating the \ac{GUBM}-Zshape in terms of behavior prediction and relevance estimation, we also include some case studies to gain a better understanding of its advantages and limitations. 
Three examples of image search results are shown in Fig.~\ref{case_study}. 
We present the top re-ranking results of our model on the left and the results of the original ranking on the right. 
Red and purple boxes mark images with a low topical relevance score and poor image quality, respectively, according to the judgments detailed in Section~\ref{subsection:datasets}. 
In the top two cases (for the queries ``loving heart'' and ``banana''), our model provides better result rankings in terms of NDCG. 
In the original ranking of ``loving heart,'' the first and  third results are not relevant to the query, which can easily be noticed as ``heart'' is usually a simple visual object in images. 
Thus, the ranking of these images with just a few interactions will be lowered. 

In the second case, our model works well on moving results with observable quality flaws to a lower ranking, especially images with big watermarks that cover major parts of the main objects. However, we find that the fourth result in the original ranking is moved to the first position in the re-ranked list produced by \ac{GUBM}-Zshape. Although this image has no obvious flaws, it has no artistic value either and does not seem appealing at all. A possible reason is that this image differs from other surrounding images, especially its color, and attracts the attention of users, which can be explained by the appearance bias~\cite{liu2015influence,wang2013incorporating}. 

Appearance bias may also affect the performance of our model in the third case (``Michael Jordan'', the worst result in our annotated dataset). 
We can observe that our model re-ranks images with low topical relevance score to the top positions. 
Although the subjects of these images are similar to the query as they all depict basketball players, their contents cannot be fully described by the query, which leads to low topical relevance level. 
The appearance difference here can be the number of objects in the image. 

To sum up, user interaction is valuable for result ranking as it records user preferences but it is affected by appearance bias, which should be paid attention to. We leave the investigation of how appearance bias affects user behavior and how to alleviate this bias for future work.


\section{Related work}
\label{section:related_work}

User interaction behavior, especially click information, has long demonstrated its value when used to improve the ranking of general Web search engines. 
By following certain examination assumptions, click models~\cite{dupret2008user,wang2015incorporating,chapelle2009dynamic} can simulate user behavior and achieve promising performance in alleviating position bias, click prediction and relevance estimation. 
Click through data has also been used as a relevance signal to train learning to match or learning to rank models~\cite{joachims2002optimizing,agichtein2006improving,hofmann-reusing-2013} and as a signal for evaluation~\citep{hofmann-online-2016}. 

Besides click information, other types of user behavior such as cursor movement, hovering, mouse scrolling and dwell time are shown to be effective in improving the performance of searcher models~\cite{shapira2006study,guo2008exploring,huang2012improving,borisov-click-2018}. Specifically, \citet{huang2012improving} demonstrate that the searcher model with the additional cursor data (i.e., hovering and scrolling) can better predict the future clicks. 

The above publications focus on general Web search. In the case of multimedia search there are many studies into \emph{why} people search for multimedia~\citep[see, e.g.,][for a recent sample]{xie-why-2018}, but there exist relatively few publications on investigating how to leverage interaction behavior to improve the performance of multimedia search engines (e.g., image or video search). Previous work~\cite{pu2005comparative,tjondronegoro2009study,park2015large,xie2017investigating} has illustrated that user behavior on grid-based search engine result pages differs from traditional, linear result pages. For example, image search leads to shorter queries, tends to be more exploratory, and requires more interactions compared to Web (text) search. 
How to incorporate the abundant and unique user behavior data available in a multimedia search setting still remains an open question. 
As in general Web search, in keyword-based image search, click-through data has been applied to bridge the gap between visual and textual content~\cite{jain2011learning,pan2014click,yu2015learning}. 
Specifically, \citet{yu2015learning} try to utilize visual features and click features simultaneously to obtain the ranking model. \citet{jain2011learning} use click data as a pseudo relevance signal to train a re-ranking model and also use PCA and Gaussian Process regression to address the sparsity problem of click data in image search. 
Although these models can boost the performance of search engines, they are mainly content-based without fully exploiting the ability of rich user behavior in image search. 
\citet{o2016leveraging} extract abundant user behavior features (e.g., hover-through rate) and demonstrate that combining these features with content features can yield significant improvements on relevance estimation compared to purely content-based features. 
However, to train the learning to rank framework using these features, a manually annotated dataset is needed. 

Somewhat orthogonal to the work listed above, \citet{oosterhuis-ranking-2018} consider so-called complex ranking settings where it is not clear what should be displayed, that is, what the relevant items are, and, importantly, how they should be displayed, that is, where the most relevant items should be placed; this is  relevant when different display formats beyond a ranked list or grid are being used for result presentation.
The authors introduce a deep reinforcement learning method capable of learning both the layout and the best ranking given the layout, from weak reward signals.

Inspired by click models developed for general Web search, we utilize user interactions on image \acp{SERP} to construct a ranking model. 
We simulate user examination behavior between adjacent interaction signals and estimate internal parameters including user examination and query-image pair relevance from observable user interactions (click and hover). 
In comparison with previous models in image search, our model is interaction-based without having to incorporate content features. 
Also, unlike the supervised learning to rank framework, our model can be trained on large-scale commercial search logs without annotated data. 
To the best of our knowledge, this is the first work to integrate grid-based user examination behavior into a user model of image search scenario.


\section{CONCLUSION AND FUTURE WORK}
\label{section:conclusionandfuturework}

We have conducted exploratory analyses to investigate the correlation between interaction signals (click and hover together) and user examination in Web image search. 
By answering three research questions, we find that cursor hovering can be an additional signal for relevance and users tend to examine results in one direction with possible skips both in the vertical direction and in the horizontal direction.
We then propose three assumptions to characterize user interaction behavior in an image search scenario. 
Based on these assumptions, we construct a interaction behavior model, called grid-based user browsing model~(\ac{GUBM}). \ac{GUBM} utilizes observable data (interaction sequences) to estimate internal parameters including the examination probability and relevance level of a given query-image pair. 

Through extensive experimentation, we demonstrate that in terms of behavior prediction and result ranking, \ac{GUBM} significantly outperforms the state-of-the-art baseline models as well as the original ranking of a commercial image search engine. 
We show that the proposed assumptions are close to actual user behavior and that integrating cursor hovering can highly improve model performance both in topical relevance and image quality. 
As \ac{GUBM} is based on interaction features on image \acp{SERP}, it can easily be transferred to other search environments that have grid-based result interfaces such as video search or product search.

Through case studies of several queries, we obtained a better understandings about the advantages and the limitations of \ac{GUBM}.
The limitations guide interesting directions for future work: (1)~The so-called appearance bias affects user interaction which may lead the behavior models to worse performance. 
How to alleviate appearance bias in multimedia search engines deserves more investigation. 
(2)~As dwell time may also be a valuable and informative signal for user preference~\citep{borisov-context-aware-2016}, we will try to model the temporal information of user interactions in the future.

\subsection*{Code and data}
To facilitate reproducibility of our results, we share the code and data used to run our experiments at \url{https://tinyurl.com/yb3etwm8}.

\subsection*{Acknowlegdements}
This research was supported by
the Natural Science Foundation of China (Grant No. 61622208, 61732008, 61472206),
the National Key Basic Research Program (2015CB358700),
Ahold Delhaize,
Amsterdam Data Science,
the Bloomberg Research Grant program,
the China Scholarship Council,
the Criteo Faculty Research Award program,
Elsevier,
the European Community's Seventh Framework Programme (FP7/2007-2013) under
grant agreement nr 312827 (VOX-Pol),
the Google Faculty Research Awards program,
the Microsoft Research Ph.D.\ program,
the Netherlands Institute for Sound and Vision,
the Netherlands Organisation for Scientific Research (NWO)
under pro\-ject nrs
CI-14-25, 
652.\-002.\-001, 
612.\-001.\-551, 
652.\-001.\-003, 
and
Yandex.
All content represents the opinion of the authors, which is not necessarily shared or endorsed by their respective employers and/or sponsors.

\bibliographystyle{ACM-Reference-Format}
\bibliography{sigir2017} 

\end{document}